\documentclass[11pt]{article}
\usepackage{moriond,epsfig}

\def\NPB{{\em Nucl. Phys.} B}
\def\PLB{{\em Phys. Lett.}  B}
\def\PRL{{\em Phys. Rev. Lett.} }
\def\PRD{{\em Phys. Rev.} D}

\newcommand{\dz}{D\O~}

\def\be{\begin{equation}}
\def\ee{\end{equation}}
\def\bea{\begin{eqnarray}}
\def\eea{\end{eqnarray}}

\begin{document}
\vspace*{4cm}
\title{$b$-production at the Tevatron}

\author{ Eric Kajfasz for the CDF and \dz collaborations}

\address{Fermilab, P.O. Box 500, Batavia, IL 60510, USA\\
CPPM, Case 907, Facult\'e des Sciences de Luminy, 13288 Marseille Cedex 9, France}

\maketitle\abstracts{
The phenomenology of $b$-production and some of the previous Run I measurements of CDF and \dz 
are briefly reviewed. A new analysis by CDF of Run I $b\bar{b}$ angular correlations and a 
new measurement by \dz of Run II $b$-jet cross-section, are presented.}

\section{Introduction}
The measurement of $b$ quark production in high energy hadronic collisions 
provides an essential test bench for how well we understand QCD,
especially in its perturbative regime. Indeed, the $b$ quark is heavy 
enough (m$_b\gg \Lambda_{QCD}$)
to justify perturbative expansions,
however, it is still light enough to be produced copiously at the Tevatron.\\
After summarizing some of CDF and \dz Run I results, 
we present the new preliminary results on B-hadron correlations that CDF recently extracted 
from its Run I data, and a preliminary \dz $b$-jet cross-section coming from an analysis 
of Run II data collected with the upgraded \dz detector. 

\section{Some of previous CDF and \dz Run I results}
\label{sec:runIres}
\begin{figure}[th]
\centering
\begin{minipage}{3in}
\centering
\epsfig{file=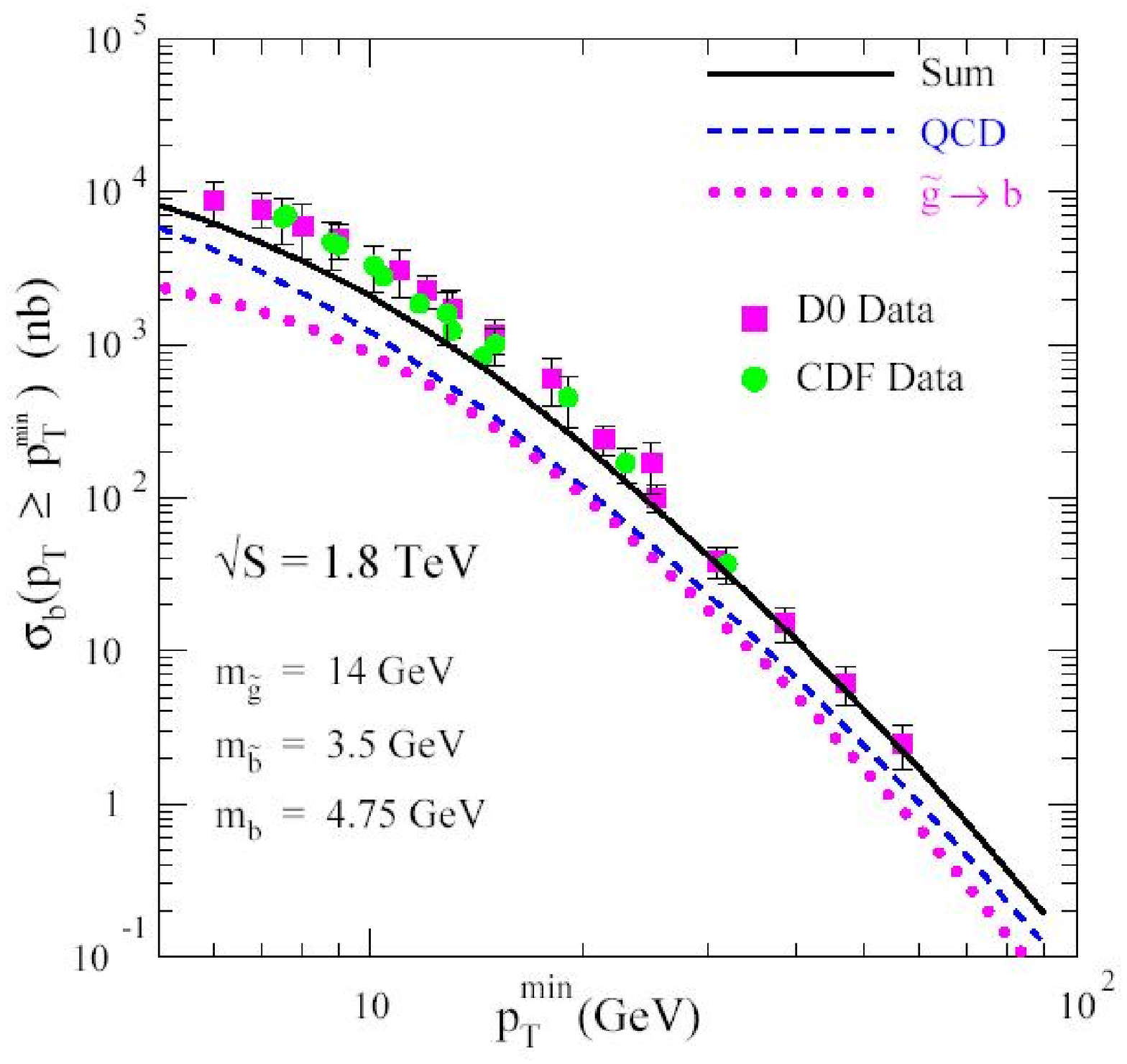,width=3in}
\caption{CDF and D0 Run I data compared to NLO QCD predictions 
and to the model discussed in $^5$.}
\label{fig:sullivan}
\end{minipage}
\begin{minipage}{3in}
\centering
\epsfig{file=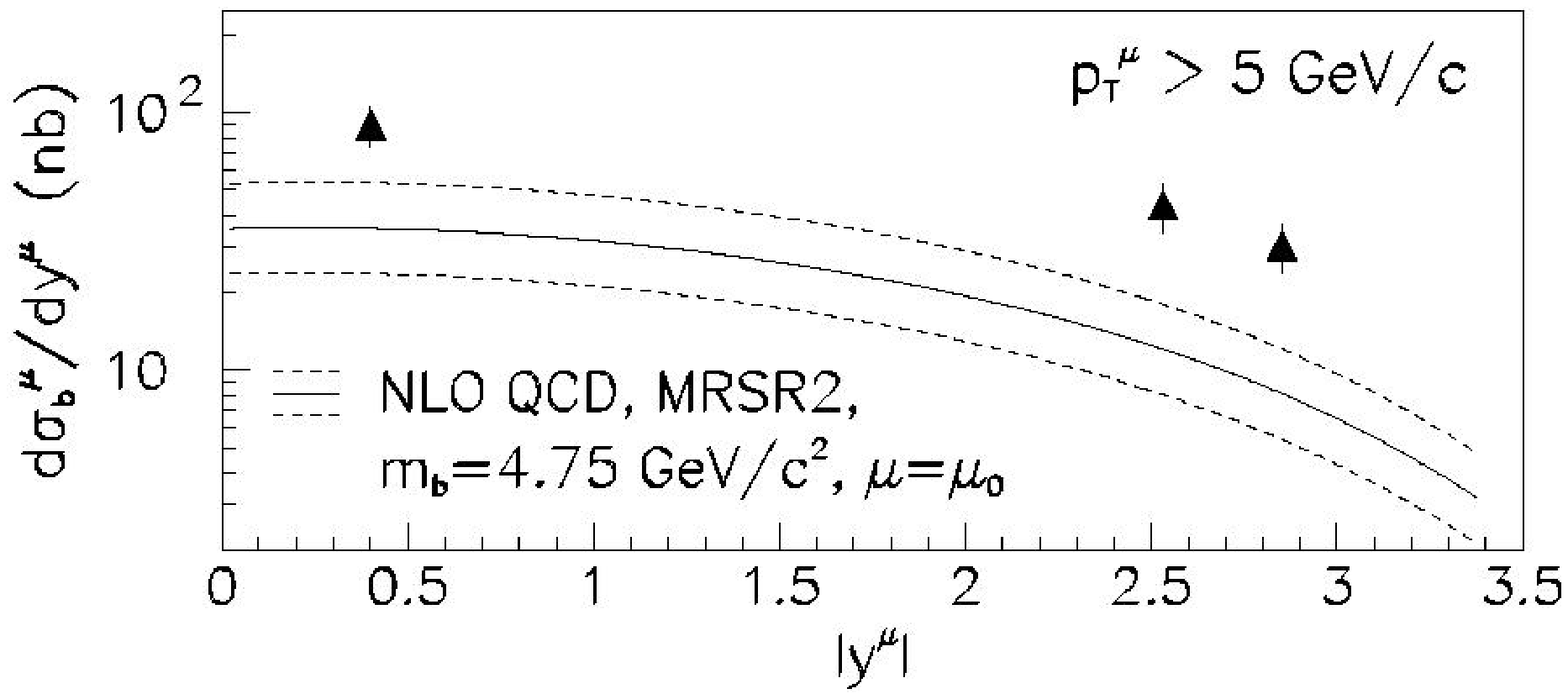,width=3in}
\caption{D0 Run I measurement of $b$ production in the forward region compared to
NLO QCD predictions.}
\label{fig:d0-3}
\end{minipage}
\end{figure}

Theoretical spectra at Next to Leading Order (NLO) have been available 
for some time \cite{nlo}, but they show a discrepancy 
by up to a factor 2 to 4 
with respect to experimental spectra measured by CDF \cite{CDFRI} and \dz \cite{D0RI} in 
Run I at the Tevatron.\\
Fig.~\ref{fig:sullivan} shows a compilation of CDF and \dz measurements 
of the integrated $b$-quark cross-section as a function of the minimum transverse momentum of 
the $b$, $P_T^{min}$, in the central rapidity region $|y^b|<1$. The measurements are compared to 
the NLO QCD prediction shown as a dashed line. 
As a general trend, this discrepancy seems to be less accute at higher $P_T^{min}$.
\dz also showed \cite{D0RI} (see fig.~\ref{fig:d0-3}) that this discrepancy worsens 
at higher rapidities.
\newpage

\begin{figure}[h]
\begin{minipage}{3in}
\centering
\epsfig{file=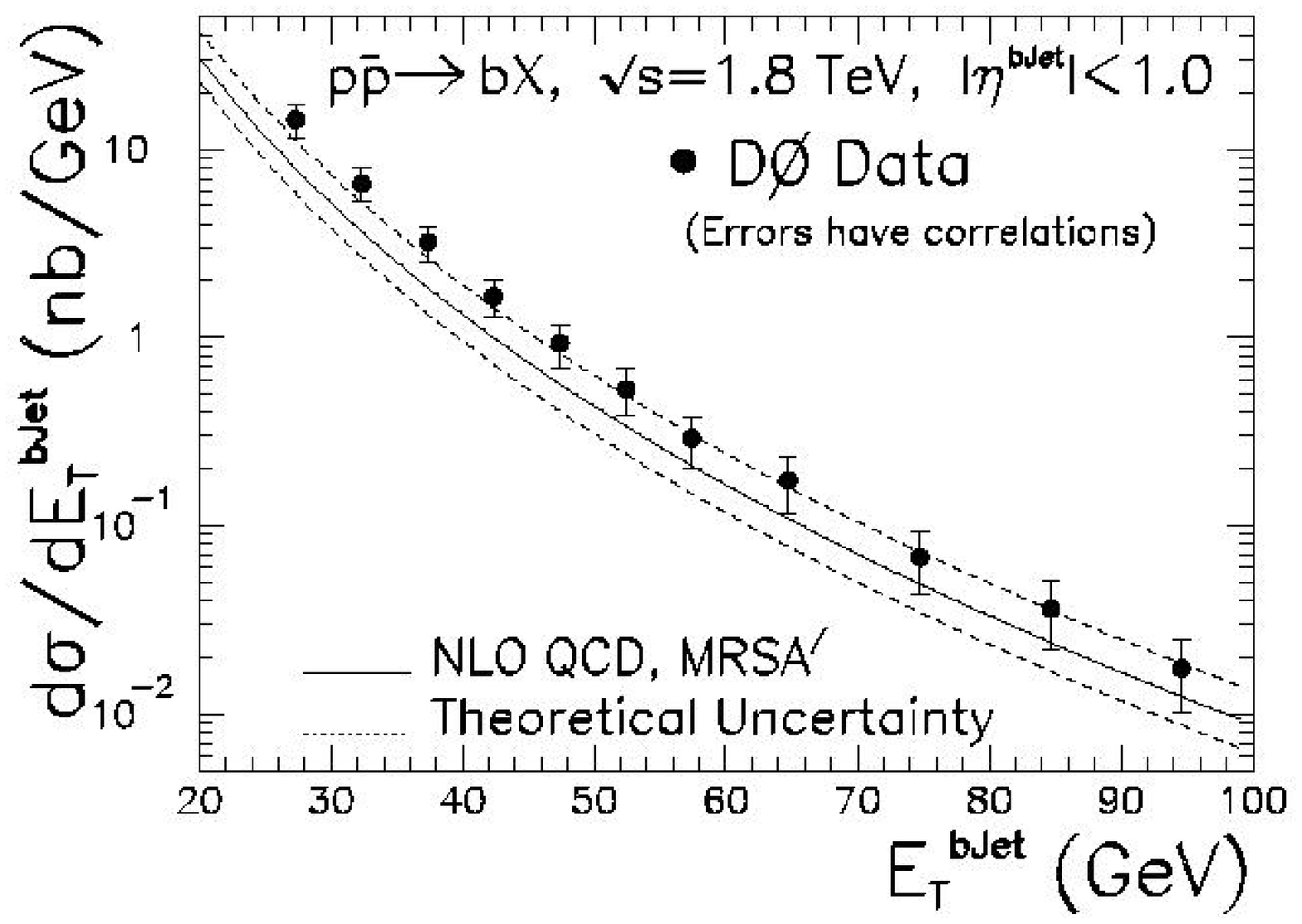,width=3in}
\caption{D0 Run I B-hadron production cross-section compared to NLO QCD predictions.}
\label{fig:bjet-runI}
\end{minipage}
\begin{minipage}{3in}
\centering
\epsfig{file=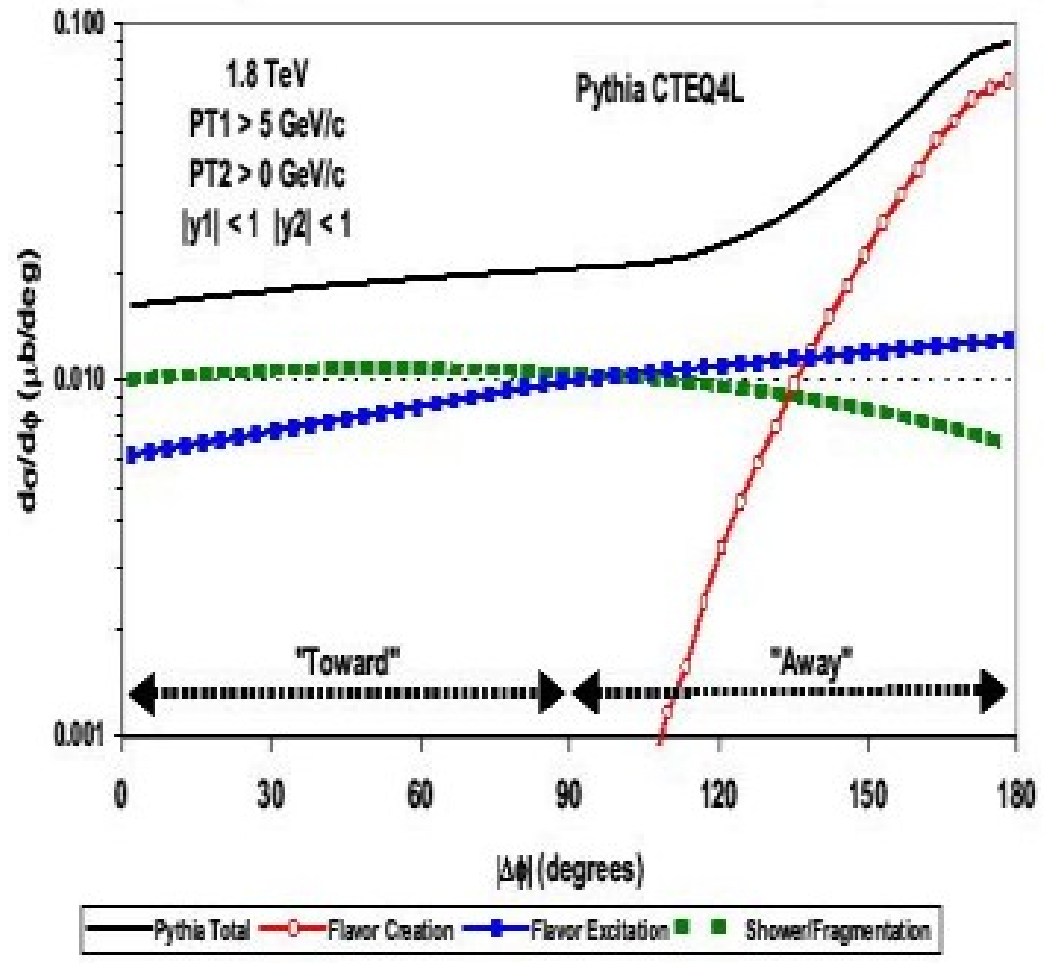,width=3in}
\caption{Predicted PYTHIA angular correlations for the 3 sources of beauty (from $^6$).
$\Delta\phi$ is the azimuthal angle between $b$ and $\bar b$.}
\label{fig:bhad-corr-pred}
\end{minipage}
\end{figure}

\noindent $P_T$ distributions for 
open $b$-quark production are sensitive to large log terms, 
which need to be resummed 
to all orders, and to non-perturbative corrections required to account for the fact 
that hadrons, not quarks, are the measured final states. To prevent such a sensitivity,
it has been suggested \cite{mangano} 
that one could look at $E_T$ distributions 
of $b$-jets, instead. Fig.~\ref{fig:bjet-runI} shows the measurement performed by D0 in run~I 
compared to the NLO QCD calculations \cite{mangano}. The better agreement
between theory and measurement seen here may hint at some possible improvement to be made 
e.g. in the heavy quark fragmentation functions.

\begin{figure}[h]
\begin{minipage}{6in}
\centering
\epsfig{file=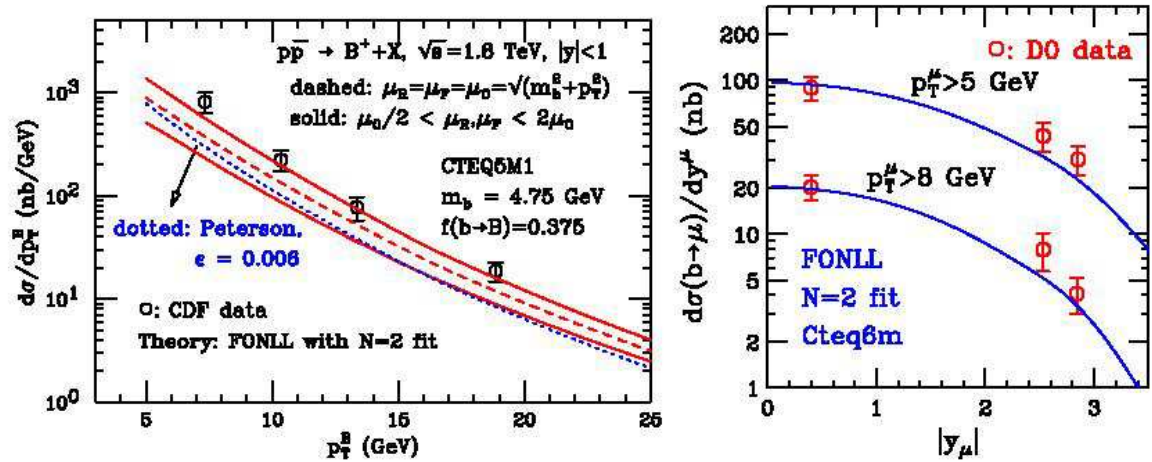,width=6in}
\caption{Run I data using tweaked resummation and fragmentation functions as discussed in $^7$.}
\label{fig:frag}
\end{minipage}
\end{figure}

\noindent The work done to try to better understand the situation mainly goes into three directions:
\newline
{\em Sources of $b$-quarks}: at leading order, $b$-quarks are produced through $q\bar q$ annihilations 
and $gg$ fusion. But, at higher orders, two other mechanisms come into play, namely, 
flavor excitation
and gluon splitting \cite{field}. The amount with which these different production
mechanisms contribute at the Tevatron depends on a number of theoretical uncertainties.
From an experimental point of view, as shown in Fig.~\ref{fig:bhad-corr-pred}, the $b\bar b$ azimuthal opening angle $\Delta\phi$ 
for the three 
sources are quite different and can be used to attempt to isolate 
their individual contributions \cite{field}.
Flavor creation tends to produce $b$ and $\bar b$ mostly back to back. Its $\Delta\phi$ 
distribution does not 
show any contribution in the "Towards" region ($\Delta\phi < 90^\circ$).
This will be elaborated on in section \ref{sec:CDF}.
\newline
{\em Resummations and Fragmentation Functions}:\\
It has been shown \cite{kniehl} that NLO QCD with $\overline{\rm MS}$ renormalization scheme, 
together with non-perturbative fragmentation functions extracted from LEP and SLC data,
can give a good agreement with CDF Run I B meson cross-section measurement.\\
Using a Fixed Order plus Next to Leading Log (FONLL) $b$ quark spectrum
and fitting the moments rather than the shape of the non-perturbative fragmentation function
from $e+e-$ data, has also proven \cite{nason} to provide a better agreement
with CDF and D0 Run I measurements, as can be seen in Fig.~\ref{fig:frag}.
\newline
{\em New physics}: an interesting possibility has been explored in \cite{sullivan}.
The dotted line on Fig.~\ref{fig:sullivan} shows how the 
production ($p\bar p \rightarrow \tilde g \tilde g$) of relatively
light gluinos followed by their decay 
into $b$ and and a light $\tilde b$ could help reduce the discrepancy.

\section{New CDF $b\bar{b}$ angular correlations from Run I data}
\label{sec:CDF}
In this analysis, CDF uses 90 $pb^{-1}$ of a data sample taken during 
the run IB of the Tevatron (1994-1995) with collisions at a center of mass energy of $1.8$~TeV.
A muon or an electron trigger 
is required in order to enrich the $b$ content of the data analyzed.
Tracking information is used to reconstruct the decay vertices of both B hadrons.
Decay vertices angular correlations are then compared to PYTHIA predictions.
\subsection{secondary vertex correlations}
Figure \ref{fig:svtx-corr} shows, for the data with an electron trigger, the opening 
angle $\Delta\phi$ between the momentum 
vectors of the secondary vertices in the transverse plane. Detector effects are 
simulated in Monte Carlo. The relative contribution from flavor creation, flavor excitation,
and gluon splitting are varied to give a best match to the data. Similar distributions
are also produced for data with a muon trigger.
\subsection{measured B-hadron correlations}
\begin{figure}
\centering
\begin{minipage}{3in}
\centering
\epsfig{file=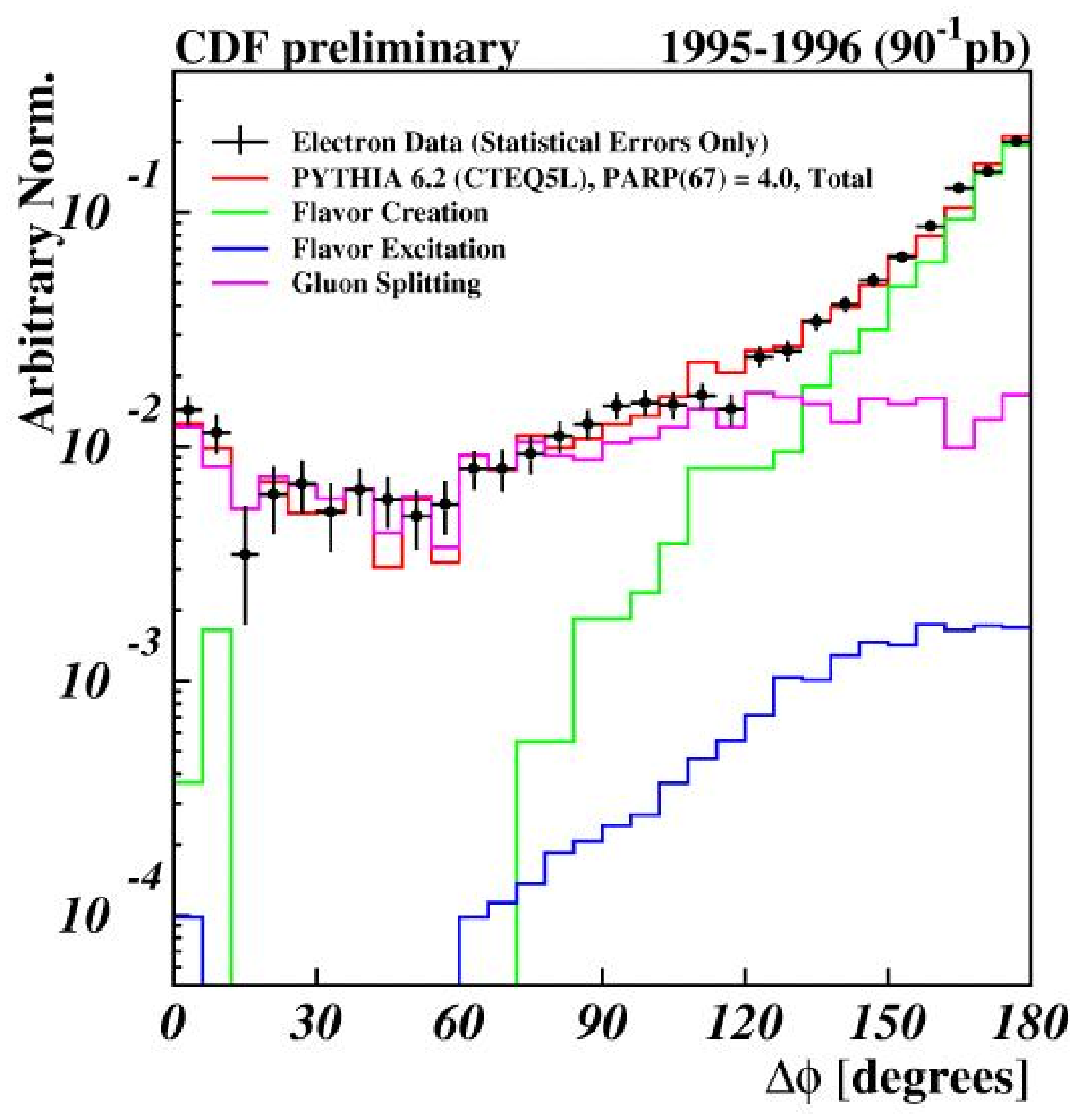,width=3in}
\caption{Measured secondary vertex correlations compared to PYTHIA with relative 
contributions of the different sources adjusted to give best match.}
\label{fig:svtx-corr}
\end{minipage}
\begin{minipage}{3in}
\centering
\epsfig{file=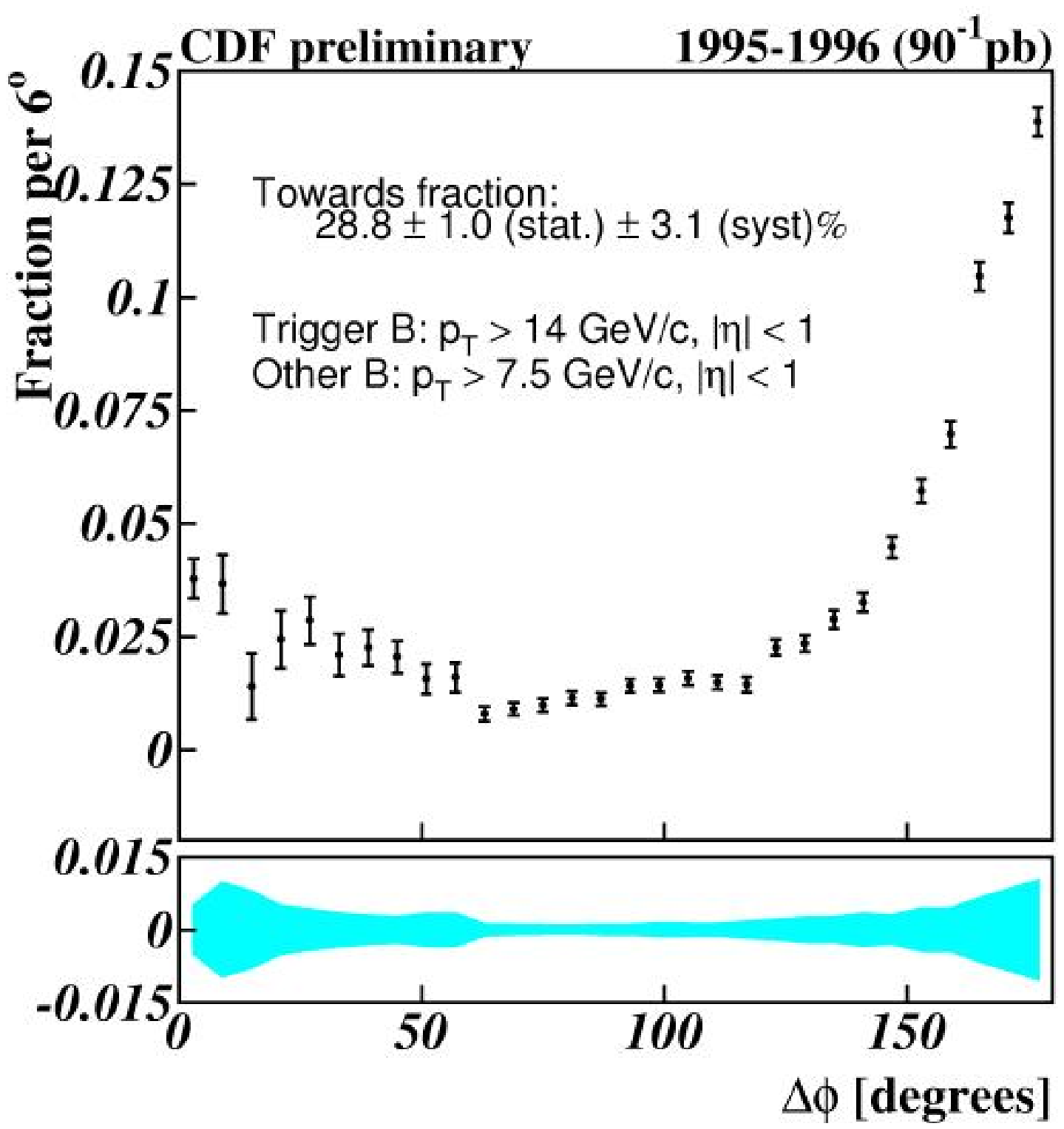,width=3in}
\caption{Measured B-hadron correlations.}
\label{fig:bhad-corr}
\end{minipage}
\end{figure}

Figure \ref{fig:bhad-corr} shows, for a total of $17,000$ e$+\mu$ events, the opening 
angle $\Delta\phi$ between the measured B directions in the transverse plane.
The detector effects are unfolded from data using PYTHIA. Also taken into account are corrections 
for mistags, tags from prompt charm, and sequential double tags.
The fraction of events in the "Towards" region ($\Delta\phi < 90^\circ$), where flavor excitation 
and gluon splitting contribute, is measured to be $28.8\pm1.0(stat)\pm3.1(syst)\%$.
The shaded region shows the correlated systematic errors.
The error bars show statistical errors only.

\section{\dz Run II $b$-jet cross-section}
This analysis utilizes 3.4 $pb^{-1}$ of data, collected with the upgraded \dz detector
between end of February and mid-May 2002 as part of the Run II of the Tevatron, with collisions 
at a center of mass energy of $1.96$~TeV.
\subsection{$\mu$+jet cross-section}
First, the cross-section for the muons associated with jets is measured.
Jets are defined with a $R=\sqrt{\Delta\eta^2+\Delta\phi^2}=0.5$ cone algorithm. The muon track is measured by the 
muon system only and the kinematic cuts used are:\\
$|\eta^{jet}|<0.6,$~~$|E_T^{jet}|>20$~GeV$,$~~$|\eta^{\mu}|<0.8,$~~$P_T^{\mu}>6$~GeV$,$~~$|\Delta R(jet,\mu)|<0.7$
\\The jet reconstruction efficiency is 100\% for $E^{jet} > 20$~GeV, the muon 
reconstruction efficiency is $43.7\pm0.8(stat)\pm2.2(syst)\%$.
\subsection{$b$-tagging and $b$-jet fraction}
\begin{figure}[h]
\centering
\epsfig{file=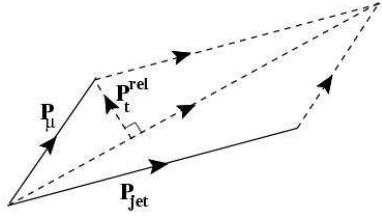,width=2in}
\caption{Definition of $P_T^{rel}$}
\label{fig:ptrel-def}
\end{figure}

Then, the $b$ content of the sample is estimated by identifying the jets emanating
from $b$ quarks. In this analysis, the tagging of $b$-jets is done by calculating the
$P_T^{rel}$ for the muons associated with jets 
(see $P_T^{rel}$ definition in Fig.~\ref{fig:ptrel-def}).
The method is based on the fact that, because of 
the high mass of the $b$ quark, the muons
produced in the decay of the $b$-quark have a higher $P_T^{rel}$ than the muons produced 
in other processes.

\noindent Since the shape of the $P_T^{rel}$ distribution depends on the energy of the jet, 
the $E_T^{jet}$ range is divided into several bins. In each of these bins, the $P_T^{rel}$
distribution for the data is adjusted to the sum of a signal template 
(extracted from a $b\rightarrow\mu$ Monte Carlo simulation) and of a background template
(extracted from 1.5 million QCD events). Because of the satistical limitations of the
background templates, four $E_T^{jet}$ bins are used. The resulting $b$-jet fraction as a 
function of $E_T^{jet}$ is given in Fig.~\ref{fig:bcont}, including a functional form 
$a+b/E_T^{jet}$ fitted to the measurements.

\subsection{$b$-jet cross-section}
\begin{figure}[h]
\centering
\begin{minipage}{3in}
\centering
\epsfig{file=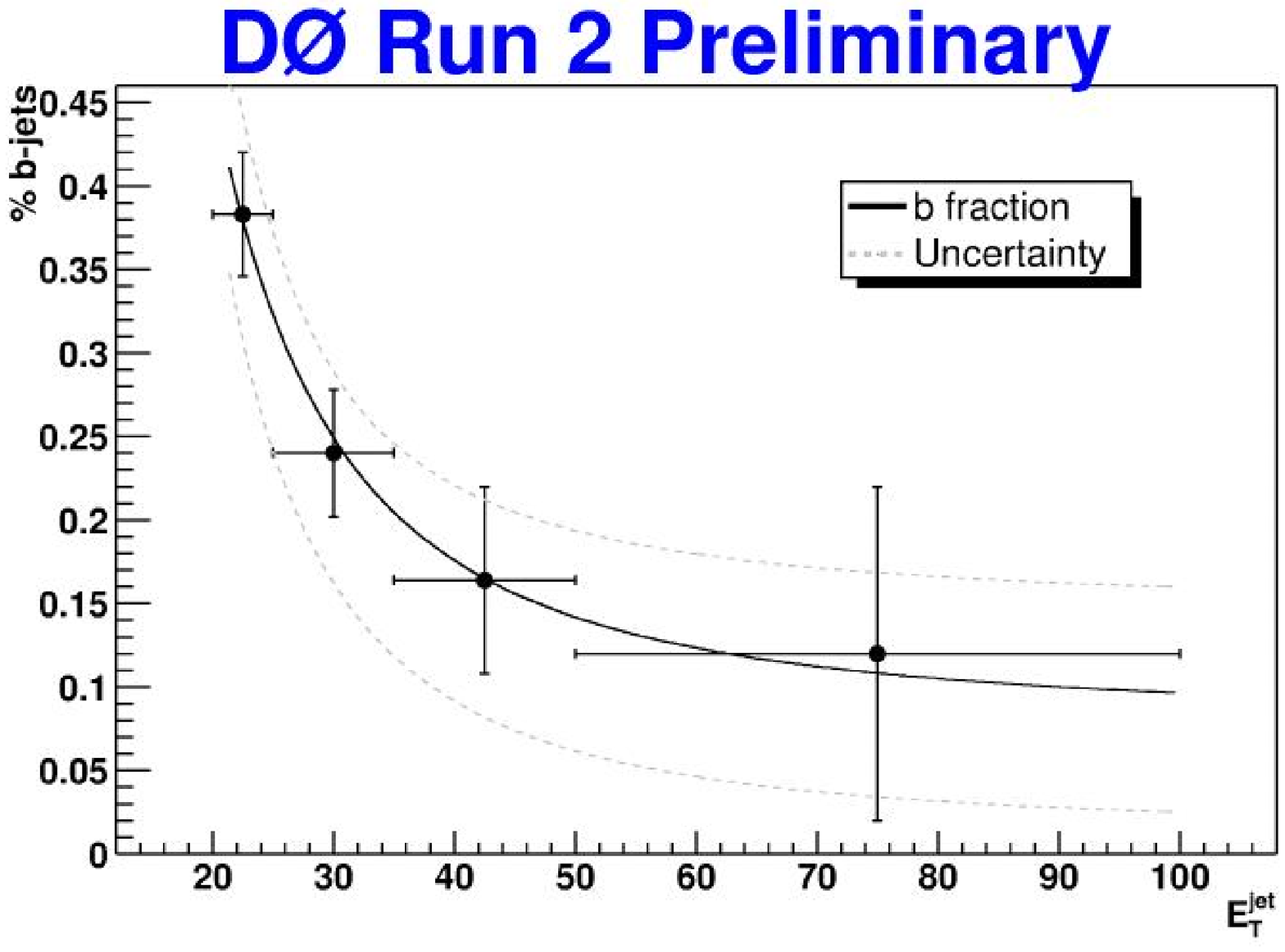,width=3in}
\caption{$b$-jet fraction as a function of $E_T^{jet}$.}
\label{fig:bcont}
\end{minipage}
~~~
\begin{minipage}{3in}
\centering
\epsfig{file=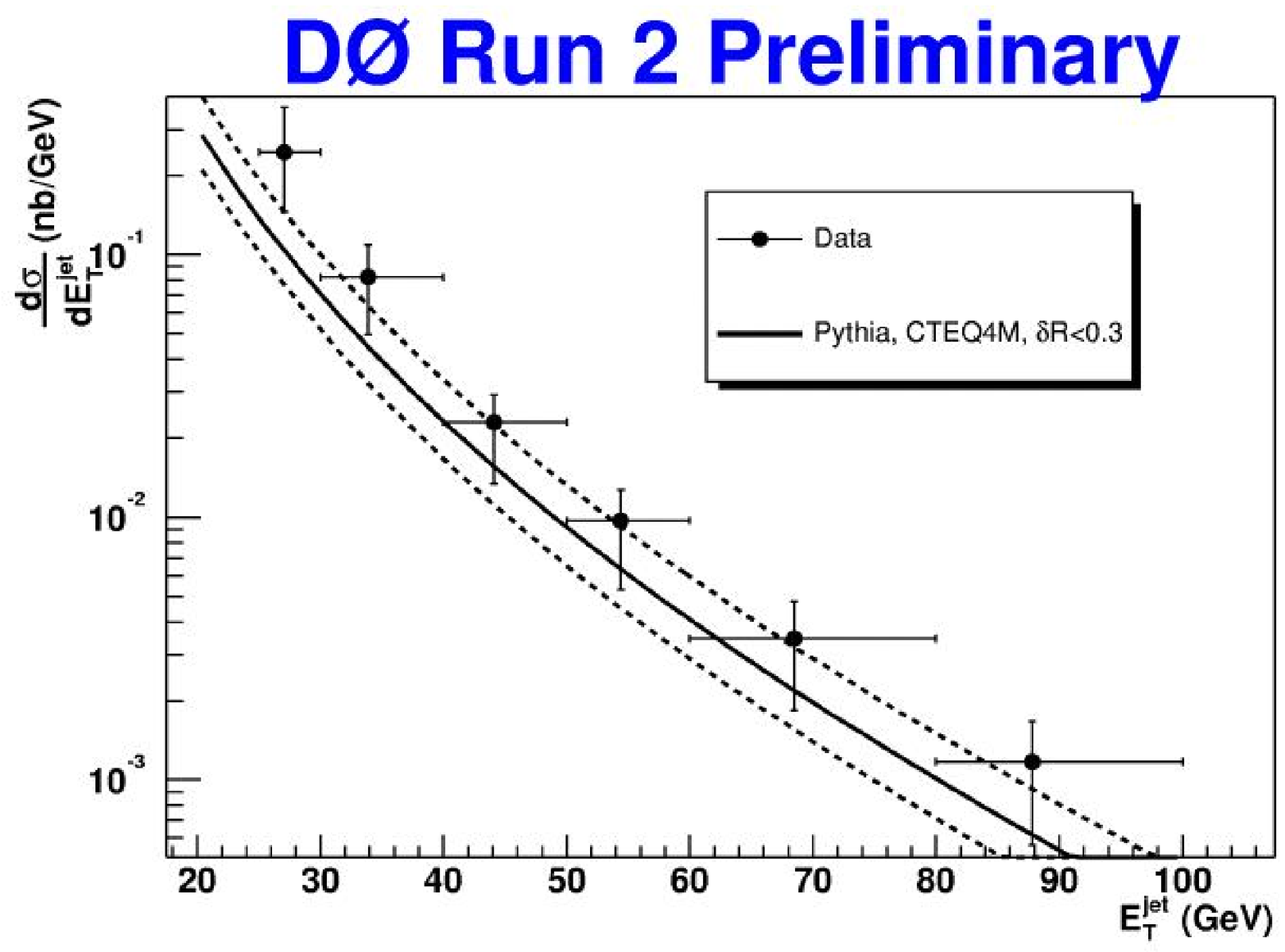,width=3in}
\caption{Measured Run II b-jet cross section compared to theoretical predictions.}
\label{fig:bjetcross}
\end{minipage}
\end{figure}

The $b$-jet cross-section is obtained by folding the $\mu$+jet cross-section 
with the $b$-jet fraction and by unfolding it from the calorimeter jet energy 
resolution using an ansatz function. The preliminary Run II result is shown 
in Fig.~\ref{fig:bjetcross}
compared to theoretical predictions (solid line). The band within 
the dashed lines covers the theoretical uncertainties.  
This measurement is consistent with the corresponding Run I result shown in 
section \ref{sec:runIres}.

\section{Conclusions}
Although the agreement between calculated and measured b-production spectra is still not perfect,
our understanding is steadily progressing, both on the theoretical and 
experimental sides. There is still room left for improvement and maybe new physics.
New Run II data at the Tevatron from both CDF and \dz will hopefully help to shed light 
on the remaining obscure corners of this long standing issue, soon.

\end{document}